# Semi-supervised learning towards automated segmentation of PET images with limited annotations: Application to lymphoma patients

Fereshteh Yousefirizi[1], Isaac Shiri[2], Joo Hyun O[3], Ingrid Bloise[4], Patrick Martineau[4], Don Wilson[4], François Bénard[4], Laurie H. Sehn[4], Kerry J. Savage[4], Habib Zaidi[2], Carlos F. Uribe[1,4,5], Arman Rahmim[1,4,5]

[1] Department of Integrative Oncology, BC Cancer Research Institute, Vancouver, Canada
[2] Division of Nuclear Medicine and Molecular Imaging, Geneva University Hospital, Geneva, Switzerland
[3] Seoul St. Mary's Hospital, College of Medicine, The Catholic University of Korea, Seoul, Republic of Korea
[4] BC Cancer, Vancouver, Canada
[5] Department of Radiology, University of British Columbia, Vancouver, Canada

**Abstract**

**Background:** The time-consuming task of manual segmentation challenges routine systematic quantification of disease burden, therapy response evaluation, radiation treatment planning and outcome prediction. Convolutional neural networks (CNNs) hold significant promise to reliably identify locations and boundaries of tumors from PET scans. However, the training step is commonly supervised and annotated-data hungry. Given the wide availability of unlabeled PET data, we aimed to leverage the need for high-quality annotated data via semi-supervised approaches, with application to PET images of diffuse large B-cell lymphoma (DLBCL) and primary mediastinal large B-cell lymphoma (PMBCL) managed in different centers.

**Methodology:** We analyzed $^{18}$F-FDG PET images of 292 patients PMBCL (n=104) and DLBCL (n=188) (n=232 for training and validation, and n=60 for external testing). Classical wisdom in conventional segmentation approaches (e.g. fuzzy clustering loss function (FCM)) was borrowed and adapted for use in loss function designs for deep supervised and unsupervised training. We employed FCM and MS losses for training a 3D U-Net with different levels of supervision: i) fully supervised methods with labeled FCM (LFCM) as well as Unified focal and Dice loss functions, ii) unsupervised methods with Robust FCM (RFCM) and Mumford-Shah (MS) loss functions, and iii) Semi-supervised methods based on FCM (RFCM+LFCM), as well as MS loss in combination with supervised Dice loss (MS+Dice). In addition, we studied changing the supervision level, α, within semi-supervised approaches.

**Results:** Unified loss function yielded higher Dice score (mean ± standard deviation (SD)) (0.73 ± 0.03; 95% CI, 0.67-0.8) and Jaccard index (0.58 ± 0.04; 95% CI, 0.5 – 0.66) compared to Dice loss (p-value<0.01). Semi-supervised RFCM+αLFCM with α=0.3 showed the best performance, with Dice score of 0.69 ± 0.03 (95% CI, 0.45-0.77) and Jaccard index of 0.54 ± 0.04 (95% CI, 0.47-0.61), outperforming MS+αDice for any supervision level (any α) (p<0.01). The best performer among MS+αDice semi-supervised approaches with α=0.2 showed Dice score of 0.60 ± 0.08 (95% CI, 0.44-0.76) and Jaccard index of 0.43 ± 0.08 (95% CI, 0.27-0.59) compared to other supervision level in this semi-supervised approach (p<0.01).

**Conclusion:** Semi-supervised learning via FCM loss (RFCM+αLFCM) showed improved performance compared to supervised approaches trained by smaller amount of labeled data. Considering the time-consuming nature of expert manual delineations and intra-observer variabilities that cause inconsistent 'ground truths', semi-supervised approaches have significant potential for automated segmentation workflows.

# Introduction

Quantifying disease burden to enable improved therapy response assessment and outcome prediction are important needs in lymphoma PET scans [1]. Automated and reliable lesion segmentation from whole-body PET images is a crucial bottleneck and a prerequisite to the widespread use of quantitative imaging biomarkers in clinical practice, including radiomics analysis. This includes computation of total metabolic tumor volume (TMTV) and analyses for individual lesions.

Simple thresholding methods remain common techniques in clinical workflow despite the large heterogeneities in lesion location, size, and contrast especially in cancer types, such as lymphoma [1,2]. More advanced segmentation methods such as active contour models [3], region growing, and clustering algorithms (Gaussian mixture models (GMM) [4] or fuzzy C-means (FCM) [5]) incorporate statistical differences between uptake regions and surrounding tissues. Conventional variational segmentation approaches basically minimize an energy function such as Mumford-Shah [6] to cluster image voxels (pixels) into several classes in an unsupervised manner [6–8]. These techniques generate the pixel (voxel)-wise prediction without ground truth and supervision. Although conventional unsupervised approaches have been extensively used for medical image segmentation, they are often computationally expensive and have limited capabilities in semantic segmentation, and commonly require user input to define the needed parameters for segmentation techniques or scanner setting [9]. Prior to applying the trained AI-based model on the "unseen" new data, we should first train the model on training data. The trained model will perform well on the "unseen" data if the training process was carried out effectively. On the other hand, conventional segmentation techniques should be applied on each scan so that they can be more reliable for segmenting "unseen" images, though they may not be computationally efficient.

Recent segmentation approaches are mainly focused on AI based techniques that appear disconnected from conventional approaches despite their value [10] and they don't utilized the efficiency of the conventional approaches. Deep neural networks, meanwhile, are typically trained (supervised) on a domain-specific data with limited diversity compared to its target application, hampering generalizability to "unseen" data.

The need for accurate and consistent ground truth for sufficient training data hinders application of advanced supervised learning approaches for tumor segmentation in PET scans. Ground truth in medical image segmentation is defined as the actual boundary of the object of interest that should be determined by histopathological analysis of an excised tumor. By this definition, ground truth is not always available [11]. The consensus of several manual segmentation by different experts is the next best alternative as ground truth (and in practice often a single expert delineates a given tumor) [12,13]. Meanwhile, intra- and inter-observer variabilities render ground truths less reproducible [14] and these limitations are reflected in supervised learning approaches. As such, unsupervised segmentation approaches can help reduce the effect of the uncertainty and inconsistency of ground truths during the learning process. The challenges of data-hungry supervised AI techniques for segmentation include: i) lack of consensus on ground truth generation; ii) time-consuming task of manual labeling that suffers from intra- and inter-observer and center variabilities; iii) lack of precision at the edges of predicted masks that can be caused by imprecise GTs (due to ground truth inconsistency) and loss function definitions that are irrelevant to the task.

The performance of AI-based techniques improves logarithmically with the training data size [15,16]. It has been shown that the performance of the AI-based approaches will not necessarily improve by increasing the number of cases for training [17]. An explanation for this is the issue of ground truth consistency. Meanwhile, the optimum number of labeled data to train the data-hungry deep learning approaches is not straightforward to know. Some studies do not differentiate this concept in deep learning and machine learning approaches and discussed that increasing the amount of labeled data will no longer improve the AI approach performance [18] which is not always the case for deep learning approaches.

These limitations motivate the use of advanced AI techniques that can be trained with different levels of supervision to alleviate the manual task of ground truth generation (labeling, annotation). Different levels of supervision can be used for training a segmentation model, from pixel/voxel-level annotations in supervised learning [19–21], image-level or inaccurate annotations in weakly-supervised learning [22], to no annotations in unsupervised learning [23] [24–26]. A fully AI based unsupervised segmentation framework for tumor segmentation in PET and PET/CT is still lacking. Although self-supervised pre-training approaches have been shown to improve label efficiency across a variety of medical tasks, they still necessitate a supervised fine-tuning step [27]. Lian et al. [28] proposed an unsupervised technique for automatic segmentation of 3D PET-CT images using a belief function to model the uncertain image information and an adaptive distance metric to consider the spatial information. The joint unsupervised learning was suggested as a segmentation approach that progressively clusters images and learns deep representations using a convolutional neural network. The combination of joint unsupervised learning and clustering such as k-means was also suggested for medical image segmentation [25].

Loss function defines the optimization problem and directly affects the segmentation model convergence during training. The role of the loss function is critical in the deep learning pipeline. The loss functions are defined based on intensity clustering with spatial information consideration. Kim et al. [29] proposed a loss function based on Mumford-Shah (MS) functional that can be used for unsupervised (self-supervised) segmentation (self-supervised and unsupervised learning techniques are sometimes used interchangeably in the literature). They showed that discrete implementation of MS loss function can be considered as k-means clustering with total variational regularization that suppresses noise in the membership function [29]. MS loss was used to train the segmentation network without the need to ground truth labels or even weak bounding box annotations. MS loss term also can be added to the dice or cross entropy losses as a data-adaptive regularized function in supervised approaches to help the network to improve the segmentation performance [29]. Considering the inherent suitability of fuzzy clustering to nuclear medicine's low-resolution characteristics, loss functions based on FCM was recently suggested [30] that can integrate fuzzy clustering and deep learning approaches for supervised, semi-supervised and un-supervised segmentation.

AI has the potential to quantify the disease burden by identifying the location of the abnormality and segmenting the lymphoma lesions [1,17,31–33]. In this paper, we considered the performance of semi-supervised approaches for lymphoma lesion segmentation in PET scans of DLBCL and PMBCL patients. Specifically, our contributions are three-fold; i) Investigating the effectiveness of different semi-supervised approaches. ii) Evaluating the capability of semi-supervised approaches for multiple tumor segmentations in lymphoma PET scans. iii) Considering the effect of ground truth consistency on the segmentation performance by adjusting the supervision level. In what follows, are methods are elaborated, followed by results, discussion and conclusions.

## Methods

### PET scans

Table 1 summarizes the data used in this study. PET images (n=292) included baseline and interim scans of patients with DLBCL (n=90) from two different centers (BC Cancer (BCC) (n=86) (limited stage diagnosed after 2005) and St. Mary's Hospital (SM) (n=102) (diagnosed after 2014 and stages varied from I to IV), and patients with PMBCL (n=104) from BC Cancer.

**Table 1.** Multi-center dataset information from different lymphoma types.

| Center | Lymphoma type | Matrix size | Voxel spacing (mm$^3$) | Average Injected Radioactivity (MBq) | Scanner Models |
|---|---|---|---|---|---|
| BC Cancer, Canada (BCC) | PMBCL | 168×168 (n=20) 192×192 (n=84) | 4.06×4.06×2 (n=20) 4.06×4.06×3.27 (n=119) | 347.5±52.6 | GE (Discovery D600 and D690) |
| BC Cancer, Canada (BCC) | DLBCL | 192×192 (n=86) | 3.65×3.65×3.27 (n=86) | 335.9±50.8 | GE (Discovery D600 and D690) |
| St. Mary's Hospital, South Korea (SM) | DLBCL | 168×168 (n=27) 192×192 (n=15) | 3.65×3.65×3.27 (n=15) 3.65×3.65×5 (n=27) | 252.0±48.1 | GE (Discovery 710) |
| St. Mary's Hospital South Korea (SM) | DLBCL | 168×168 (n=60) | 4.07×4.07×5 (n=60) | 240.5±47 | Siemens (Biograph40 TruePoint) |

### Ground truth Segmentation

The ground truth volumes of interest (VOI) were delineated by experienced nuclear medicine physicians using a built in-house semi-automatic workflow for MIM (MIM Software, USA), where lesions were drawn utilizing the software's gradient-based segmentation tools (PETedge and PETedge+), designated into different body parts (neck, chest, abdomen and pelvis, muscles, bones, central nervous system and other). As previously demonstrated [34] this workflow has shown reproducibility for lesion segmentation and helps to reduce the inter-observer variability.

## Preprocessing and Data Augmentation

PET images were reshaped to a common resolution of $4 \times 4 \times 2$ mm$^3$ using linear interpolation. A slice thickness of 2 mm was chosen to retain small image details that could be lost if interpolated at a larger voxel size. PET image intensities can exhibit a high variability in both within image and between images. To reduce the intensity variabilities of PET scans, we applied Z-score normalization for each scan separately, with the mean and the standard deviation computed based only on voxels with non-zero intensities corresponding to the body region.

To increase the diversity in lesion size and shape, we utilized scaling (with a random factor) and elastic deformations for data augmentation to help the model to learn the varied size and shape of the lesions. The following augmentation techniques were applied to increase the complexity of the training data: i) Spatial and intensity transformation (i.e. rotation in random directions (< 25 degree), ii) scaling with a random factor (0.8 and 1.2), iii) elastic deformations, iv) Gamma corrections with γ sampled from the uniform distribution (0.8 and 1.2)).

## Segmentation Network Architecture

3D U-Net [35] consists of the conventional convolutional blocks composed of a 3×3×3 convolution, a normalization layer (batch norm), and a ReLU activation unction as a basic element of the network. We utilized (Figure 1) the residual blocks, supplemented with a concurrent spatial and channel squeeze and excitation module SE normalization (Fig. 1 blue blocks). The SE module gives weight to the feature maps, so that the network can emphasize its attention adaptively. The squeeze & excitation (SE) module is the architectural unit that is designed to increase the representational ability of the network by enabling dynamic channel-wise feature recalibration. SE module 'squeeze' along the spatial domain and 'excite' along the channels that helps the model to highlight the meaningful features and suppress the weak ones. We used SE normalization layers with the fixed reduction ratio (r = 2) that controls the size of the bottleneck in SE normalization layers. Besides, we switched from using batch norm layers to instance normalization to reduce the memory consumption [36].

In our network, the max pooling operations in the encoder of the network are replaced by learnable downsampling blocks (Figure 1, green blocks). The max pooling operations are replaced in the encoder of the network by learnable downsampling blocks (Figure 1, green blocks), which consist of one $3 \times 3 \times 3$ strided convolutional layer, the instance norm, the ReLU activation, and the SE module. The upsampling blocks in the decoder of the network were implemented by a 3×3×3 transposed convolution instead (Figure 1, yellow blocks).

We increased the receptive field of the network and reduced memory consumption by implementing the first downsampling block with a kernel size of 7×7×7 after the input. The sigmoid activation function after the last convolutional layer produces the model output with a kernel size of 1×1×1. Due to the large size of PET images, we trained the 3D model on randomly extracted patches of $128 \times 128 \times 64$ voxels with a batch size of 2.

We trained the model for 400 epochs using Adam optimizer with β1 = 0.9 and β2 = 0.99 for exponential decay rates for moment estimates. We applied a cosine annealing schedule, gradually reducing the learning rate from $lr_{max} = 10^{-4}$ to $lr_{min} = 10^{-6}$ for every 25 epochs with adjustment at each epoch. Adam optimizer [41] was used to train all models. All models were implemented using Python with *PyTorch* library. We trained and tested all models on NVIDIA V100 GPUs.

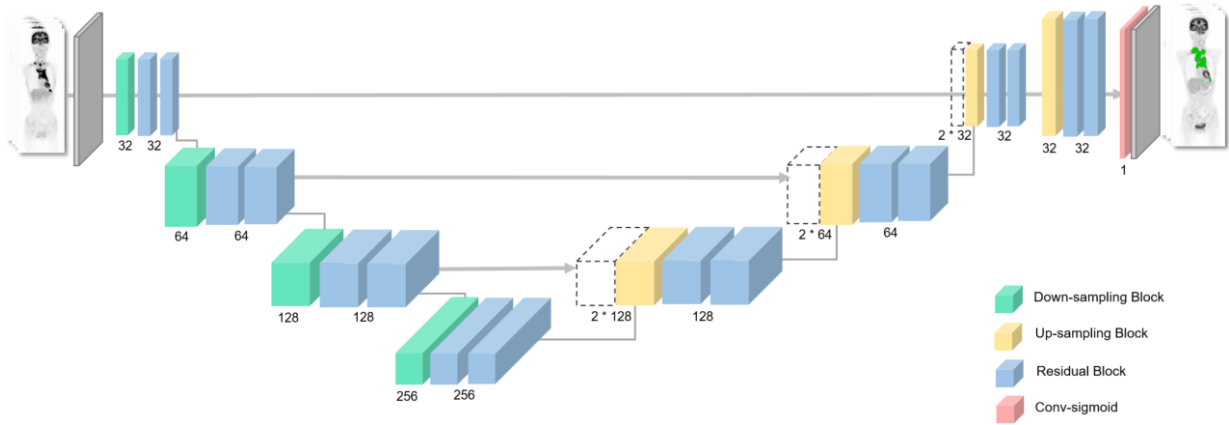

**Figure 1.** Encoder-Decoder Network with residual blocks. The number of output channels is depicted under blocks of each group. Max pooling operations in the encoder of the network are replaced by learnable downsampling blocks (green blocks). The upsampling blocks in the decoder of the network were implemented by a 3×3×3 transposed convolution instead (yellow blocks).

## Training Strategies

We applied two scenarios for training the segmentation model by semi-supervised learning. In the first experiment (experiment I), we used 104 PMBCL cases from BCC center and 188 DLBCL cases from SM and BCC centers. We trained the supervised and semi-supervised models on 60 annotated cases: 20 PMBCL cases from BCC, 20 DLBCL cases from BCC, and 20 DLBCL cases from SM. The two semi-supervised approaches used extra 172 unannotated cases (84 PMBCL cases from BCC and 22 DLBCL cases from SM and 66 DLBCL cases from BCC) for unsupervised training. We considered 60 DLBCL cases from SM center as the external test set. In the second experiment (experiment II), we evaluated two semi-supervised approaches on 122 DLBCL case: 42 cases from SM center and 80 cases from BCC. We trained the supervised and semi-supervised approaches on 30 cases: 10 from SM and 20 from BCC and considered 30 cases from BCC center as the external test set.

## Unsupervised Learning

### Fuzzy clustering-based loss functions

Unsupervised learning techniques help tackle the fact that supervised techniques can be biased given the commonly limited size and diversity of labeled data in medical imaging. The classical wisdom in conventional segmentation can be borrowed and adapted to AI loss function designs for either supervised or unsupervised methods. By minimizing an objective function, that clusters the voxels, traditional clustering-based segmentation techniques like k-means, FCM, and GMM learn the statistics of the intensity information of the image. Due to its simplicity, robustness, and efficiency, FCM is a widely used clustering-based segmentation method for medical images in supervised or unsupervised learning tasks. Since spatial information is not taken into account in clustering techniques like FCM, they are prone to error in the presence of image noise and artifacts. The FCM approach has been modified in a number of ways to incorporate spatial constraints, spatial context information, and to allow the labeling of a pixel to be influenced by the labels of nearby pixels [37–42]. Robust FCM (RFCM) that has a Markov-random-field (MRF) [43] based regularization term is able to consider the membership function changes in local neighborhoods:

$$J_{RFCM} = \sum_{j\in\Omega}\sum_{k=1}^{C} u_{jk}^q \|y_j - v_k\|^2 + \beta \sum_{j\in\Omega}\sum_{k=1}^{C} u_{jk}^q \sum_{j\in N_j}\sum_{m\in M_k} u_{lm}^q \quad (1)$$

where $u_{jk}$ is the membership functions for the $j^{th}$ voxel and $k^{th}$ class. $v_k$ is the class-centroid, $y_j$ is the voxel value at location $j$, $C$ indicates the number of classes and $\Omega$ is the spatial domain of the image. The amount of fuzzy overlap between clusters is controlled by $q$ and the second term is considered as the $J_{spatial}$ in which $N_j$ defines the neighboring voxels of the voxel $j$, $M_k$ is a set containing $\{1, ..., C\} \setminus \{k\}$ (class numbers other than $k$). The weight of the spatial smoothness term is controlled by $\beta$. This optimization problem is solved using the Lagrange multiplier to enforce the constraint and applying the partial derivatives with respect to $v_k$ and $u_{jk}$. Inspired by the definition of FCM objective function, FCM loss function was suggested for supervised and unsupervised training [30].

Based on RFCM objective function, the membership functions, u are modeled by using the softmax output of the last layer ($f(y;\theta)$) [30]:

$$L_{RFCM}(y;\theta) = \sum_{j\in\Omega}\sum_{k=1}^{C} f_{jk}^q(y;\theta)\|y_j - v_k\|^2 + \beta \sum_{j\in\Omega}\sum_{k=1}^{C} f_{jk}^q(y;\theta) \sum_{j\in N_j}\sum_{m\in M_k} f_{lm}^q(y;\theta) \quad (2)$$

where $f_{jk}(y;\theta)$ is the $k^{th}$ channel softmax output of the CNN at location $j$, and the class mean $v_k$ is defined as follows:

$$v_k = \frac{\sum_{j\in\Omega} f_{jk}^q(y;\theta) y_j}{\sum_{j\in\Omega} f_{jk}^q(y;\theta)} \quad (3)$$

### Mumford-Shah loss function

The MS loss function also helps the network utilize unlabeled images as elements of the training data.

$$L_{MS} = \sum_{k=1}^{C}\sum_{j\in\Omega} f_{jk}\|y_j - v_k\|^2 + \eta \sum_{k=1}^{C}\sum_{j\in\Omega}|\nabla f_{jk}| \quad (4)$$

where $f_{jk}$ is the softmax output of CNN, considering that $\sum_{k=1}^{C} f_{jk} = 1$ and $|\nabla f_{jk}|$ is the approximation of total variant of $f_{jk}$ and can be approximated by $\nabla f_{jk} = f_{(j+1)k} - f_{jk}$. The average voxel intensity is shown by $v_k$ here as well. By considering $\eta = 0$, the $L_{MS}$ loss function is changed to a FCM loss with $q = 1$ [30].

### Semi-supervised Learning

Incorporating a weighted supervised loss and the unsupervised loss is the basic idea of the semi-supervised learning techniques that is firstly introduced by Kim et al. [29]. They suggested incorporating a weighted supervised loss in addition to the unsupervised loss in the case of having access to labeled data:

$$L_{semi}^\alpha(y,g;\theta) = L_{unsupervised}(y;\theta) + \alpha L_{supervised}(y,g;\theta) \quad (5)$$

where α is a weighting parameter that controls the strength of the supervised term, and $g$ denotes ground truth. The parameter α is a hyperparameter that controls the weight of the supervised loss. When α is set to be small, training of the network focuses more on characterizing intensity distributions instead of the ground truth of the annotated training dataset.

### Semi-supervised learning based on Fuzzy clustering approach

Since Dice loss or cross-entropy are not immediately compatible with the 'fuzziness' of the FCM classification, a supervised loss function (Label based FCM (LFCM) [30]) based on FCM objective function can be added to design a semi-supervised loss function with fuzzy clustering based loss:

$$L_{LFCM}(y;\theta) = \sum_{j\in\Omega} \sum_{k=1}^{C} f_{jk}^{q}(y;\theta)\|g_{jk} - \mu_k\|^2 \quad (6)$$

where $g_{jk}$ is the ground truth label at location $j$ for $k^{th}$ class, and $\mu_k$ is the class mean computed within the ground truth image g and can be defined as a constant ($\mu_k = 1$). The fuzzy overlap between softmax channels is regulated by $q$.

$$L_{semi-FCM}^{\alpha}(y,g;\theta) = L_{RFCM}(y;\theta) + \alpha L_{LFCM}(y,g;\theta) \quad (7)$$

### Semi-supervised learning based on Mumford-Shah approach

Kim et al. [29] suggested to combine Mumford-Shah (MS) loss and cross entropy (CE) loss as the unsupervised and supervised losses respectively to form the semi-supervised learning approach; In this work, we tried Dice loss as the supervised term:

$$L_{semi-MS}^{\alpha}(y,g;\theta) = L_{MS}(y;\theta) + \alpha L_{Dice}(y,g;\theta) \quad (8)$$

### Supervised learning

For supervised segmentation approaches by deep neural networks, cross-entropy loss and Dice loss, or a combination of them are commonly used. The differences in segmentation performance across loss functions highlight the significance of loss function selection and how it affects the robustness and convergence of the segmentation model [17]. The three main categories of loss functions that have been used for medical image segmentation are distribution-based losses (e.g. cross entropy, Focal loss [18]), region based losses (e.g. Dice), boundary-based loss (e.g. MS [19]) and the combination of them. It has been shown that usually the best performance is observed with combined loss functions [17, 20] e.g. the sum of cross entropy and Dice similarity coefficient (DSC) [21] or the sum of Focal and Dice loss i.e. the Unified Focal loss [17].

## Quantitative Evaluation and Statistical Analysis

We evaluated the segmentation performance based on Dice similarity coefficient (DSC) and Jaccard index that are formulated as follows (considering the following definition: number of true positive (TP), Number of false positive (FP), Number of false negative (FN), Number of true negative (FN)):

$$DSC = \frac{2 \times TP}{(TP + FP) + (TP + FN)} \quad (9)$$

$$JI = \frac{TP}{TP + FP + FN} \quad (10)$$

We calculated conventional clinically relevant image-derived PET metrics, including $SUV_{max}$, $SUV_{mean}$, and $SUV_{median}$. We also extracted first-order radiomic features, including 10 and 90 percentile, energy, interquartile range, kurtosis, mean absolute deviation, rang, robust mean absolute deviation, root mean squared, total energy, and variance. All feature extractions were performed according to the image biomarker standardization initiative Using Lifex [44]. We calculated the mean relative error with respect to manual segmentation using the following relative error: Relative error (%) = ((predicted mask – ground truth)/ (ground truth))×100.

We compared the different approaches using Wilcoxon signed rank test (a non-parametric statistical hypothesis test), and reported the mean±SD and 95% confidence interval (CI)for the quantitative metrics.

## Results

Figure 2 illustrates the 2D axial views of lesion segmentation results of a patient from the external test set (SK) along with their zoomed version. The segmentations generated by the different supervised approaches are in good agreement with manual segmentations defined on lymphoma lesions presenting with different sizes, locations, textures, and contrast. The semi-supervised approach based on FCM, i.e., RFCM+αLFCM, shows better qualitative and quantitative performance than MS+αDice. Figure 3 shows the segmented lesions of the patient by the same segmentation model with different levels of supervision; the number of false positives captured by unsupervised and semi-supervised (MS+αDice) can be seen in this figure. However, unsupervised methods based on RFCM and MS losses cannot capture the lesion area correctly, as shown in Figure 3.

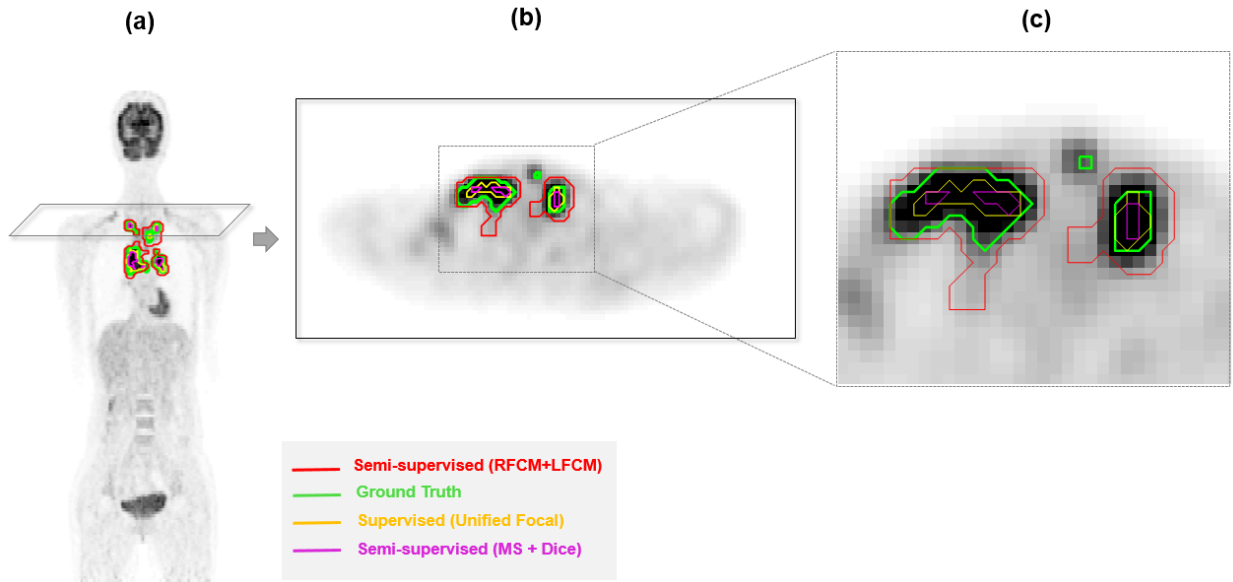

**Figure 2.** Segmentation results achieved by the different levels of supervision in supervised and semi-supervised approaches. (a) A coronal view of a DLBCL patient (b) is the axial view of the segmented lesions and (c) is the zoomed area of the segmented lesions. Visual inspection of compassion between the segmentation approaches with different supervision levels. Segmentation models with any supervision level did not segment the small lesion.

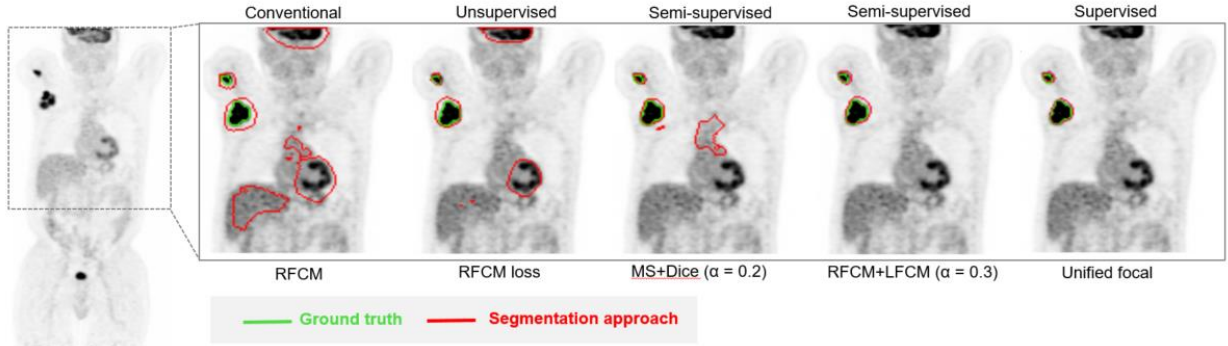

**Figure 3.** Segmentation results achieved by the different levels of supervision in unsupervised, supervised and semi-supervised AI approaches for segmentation along with the conventional segmentation based on RFCM.

Table 2 summarizes the performance of unsupervised, semi-supervised, and supervised approaches for lymphoma lesion segmentation from PET images. Figure 4 compares the Dice coefficients of the various networks to demonstrate the significance of the differences using the signed ranked test, where a p-value <0.001 is regarded as significant. The performance metrics (mean ± SD) for different methods with different levels of supervision include Dice score and Jaccard, and the lower and upper bounds are also shown. The supervised approach with Unified focal loss function yielded the highest Dice score (mean ± standard deviation (SD)) (0.73 ± 0.03; 95% CI, 0.67-0.8) and Jaccard index (0.58 ± 0.04; 95% CI, 0.5 – 0.66) compared to Dice loss (p-value <0.01). There is no significant difference between the segmentation performance of supervised Unified focal and LFCM losses. Supervised approach with label FCM (LFCM) provided the performance with Dice score of (0.71 ± 0.05; 95% CI, 0.62-0.81) and Jaccard index (0.56 ± 0.06; 95% CI, 0.55-0.68). The semi-supervised approach by RFCM and LFCM loss with α=0.3 showed the best performance among the semi-supervised approaches with Dice score (0.69 ± 0.03; 95% CI, 0.45-0.77) and Jaccard index (0.54 ± 0.04; 95% CI, 0.47-0.61) (p-value <0.01). The best performer among MS+αDice semi-supervised approaches with α=0.2 showed Dice score of (0.60 ± 0.08; 95% CI, 0.44-0.76) and Jaccard index (0.43 ± 0.08; 95% CI, 0.27-0.59) (p<0.01). It was observed that the unsupervised approach with MS loss showed the lowest performance. With the exception of few non-significant differences, as shown in Figure 4, most differences are significant and are shown in blue.

**Table 2.** Summary of Quantitative Image Segmentation Performance Metrics (Mean ± SD). Best performances in supervised, unsupervised, as well as MS+Dice and RFCM+LFCM semi-supervised methods are shown in bold.

| Methods | Training | | Hyper-parameters | External Test (n=60) | | |
|---|---|---|---|---|---|---|
| | # annotated | # unannotated | | Dice score | Jaccard | Lower-Upper Bound of 95% CI of Jaccard |
| **Supervised (Unified Focal loss)** | 232 | 0 | $\lambda=0.5, \delta=0.6, \gamma=0.5$ | **0.73 ± 0.03** | **0.58 ± 0.04** | 0.50 – 0.66 |
| **Supervised (DSC loss)** | 232 | 0 | - | 0.67 ± 0.07 | 0.51 ± 0.08 | 0.36 – 0.65 |
| **Supervised (FCM loss)** | 232 | 0 | q=2 | 0.71 ± 0.05 | 0.56 ± 0.06 | 0.55 – 0.68 |
| **Unsupervised (MS)** | 0 | 232 | $\eta=10^{-6}$ | 0.28 ± 0.17 | 0.18 ± 0.12 | 0.06 – 0.42 |
| **Unsupervised (RFCM loss)** | 0 | 232 | q=2, β=0.0016 | **0.41 ± 0.15** | **0.26 ± 0.11** | 0.04 – 0.48 |
| **Semi-supervised (MS+DSC)** | 60 | 172 | $\eta=10^{-6}, \alpha=0.1$ | 0.52 ± 0.11 | 0.36 ± 0.01 | 0.17 – 0.56 |
| | | | $\eta=10^{-6}, \alpha=0.2$ | **0.60 ± 0.08** | **0.43 ± 0.08** | 0.27 – 0.59 |
| | | | $\eta=10^{-6}, \alpha=0.3$ | 0.53 ± 0.12 | 0.37 ± 0.10 | 0.16 – 0.58 |
| | | | $\eta=10^{-6}, \alpha=0.4$ | 0.53 ± 0.12 | 0.37 ± 0.11 | 0.15 – 0.58 |
| | | | $\eta=10^{-6}, \alpha=0.5$ | 0.52 ± 0.14 | 0.37 ± 0.12 | 0.12 – 0.61 |
| | | | $\eta=10^{-6}, \alpha=0.6$ | 0.40 ± 0.14 | 0.26 ± 0.11 | 0.05 – 0.47 |
| | | | $\eta=10^{-6}, \alpha=07$ | 0.35 ± 0.14 | 0.22 ± 0.10 | 0.02 – 0.42 |
| **Semi-supervised (RFCM+FCM)** | 60 | 172 | q=2, β=0.0016, α=0.1 | 0.40 ± 0.15 | 0.43 ± 0.08 | 0.28 – 0.58 |
| | | | q=2, β=0.0016, α=0.2 | 0.61 ± 0.08 | 0.44 ± 0.08 | 0.30 – 0.60 |
| | | | q=2, β=0.0016, α=0.3 | **0.69 ± 0.03** | **0.54 ± 0.04** | 0.47 – 0.61 |
| | | | q=2, β=0.0016, α=0.4 | 0.66 ± 0.08 | 0.50 ± 0.05 | 0.39 – 0.60 |
| | | | q=2, β=0.0016, α=0.5 | 0.65 ± 0.03 | 0.48 ± 0.07 | 0.35 – 0.61 |
| | | | q=2, β=0.0016, α=0.6 | 0.60 ± 0.05 | 0.44 ± 0.06 | 0.32 – 0.55 |
| | | | q=2, β=0.0016, α=0.7 | 0.60 ± 0.07 | 0.42 ± 0.06 | 0.30 – 0.53 |

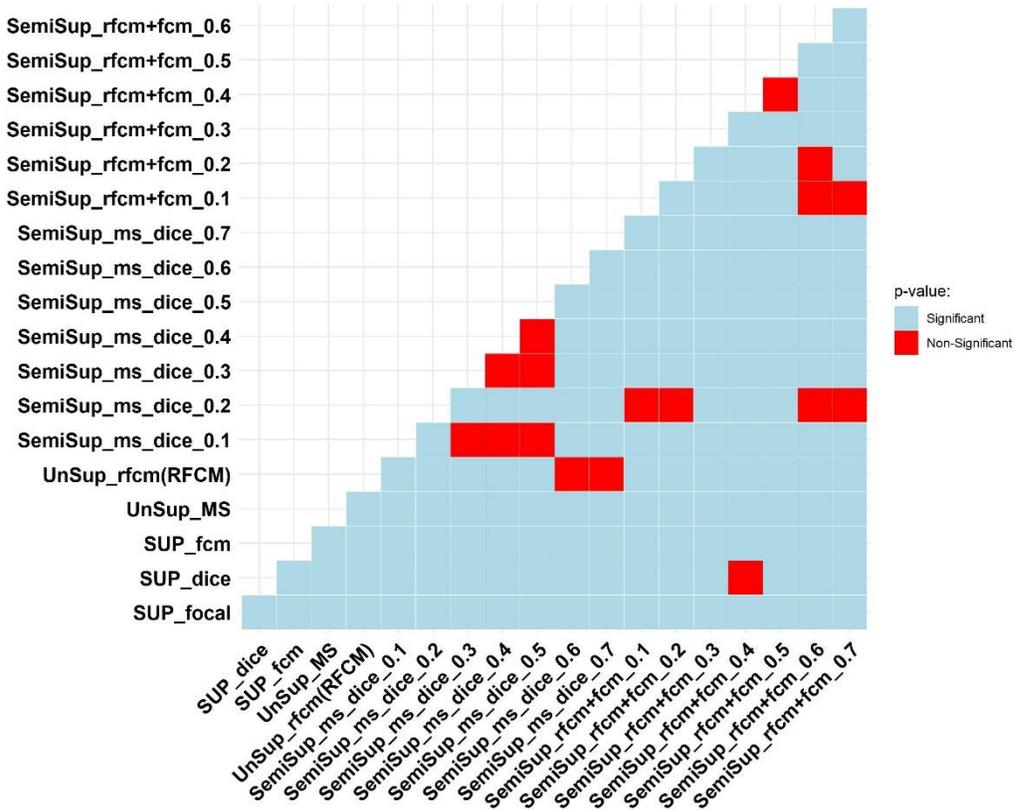

**Figure 4.** Comparison of performance of different supervision levels and loss functions (p-values) in terms of Dice coefficient (p-value<0.001 used as significant). For example we did not observe statistically significant differences between supervised approaches with Unified focal and LFCM losses, semi-supervised (MS+αDice, α=0.7) and unsupervised (MS), unsupervised (RFCM) and semi-supervised (MS+αDice, α=0.7) approaches. The significance of the different performances of the semi-supervised approaches cannot be concluded amongst all α values.

For every segmented lesion, the results of conventional image-derived PET metrics are shown in Figure 5, along with first-order (FO) and shape features percent relative error for various approaches with various levels of supervision and loss functions. This reveals that if the lesion is segmented by varying levels of supervision and losses, the segmented region contains the maximum value of the lesion, and the relative error RE for $SUV_{max}$ is less than 1%. In any case, it is seen that errors for unsupervised approach by MS loss are high. Unsupervised with RFCM and semi-supervised approaches with MS+αDice (α=0.6 and 0.7) have higher relative errors in the SUV and FO based radiomics features compared to other techniques. The percent relative error of $SUV_{mean}$ was less than 10% in techniques with some levels of supervision, including semi and supervised methods with both MS+αDice and RFCM+αLFCM losses. We also observe that with other radiomics features, errors are less than <10-20%, with the exception of FO-variance. In supervised and FCM-based semi-supervised approaches compared to unsupervised and MS-based semi-supervised techniques, the mean relative errors of SUV-based and FO-based features are relatively lower.

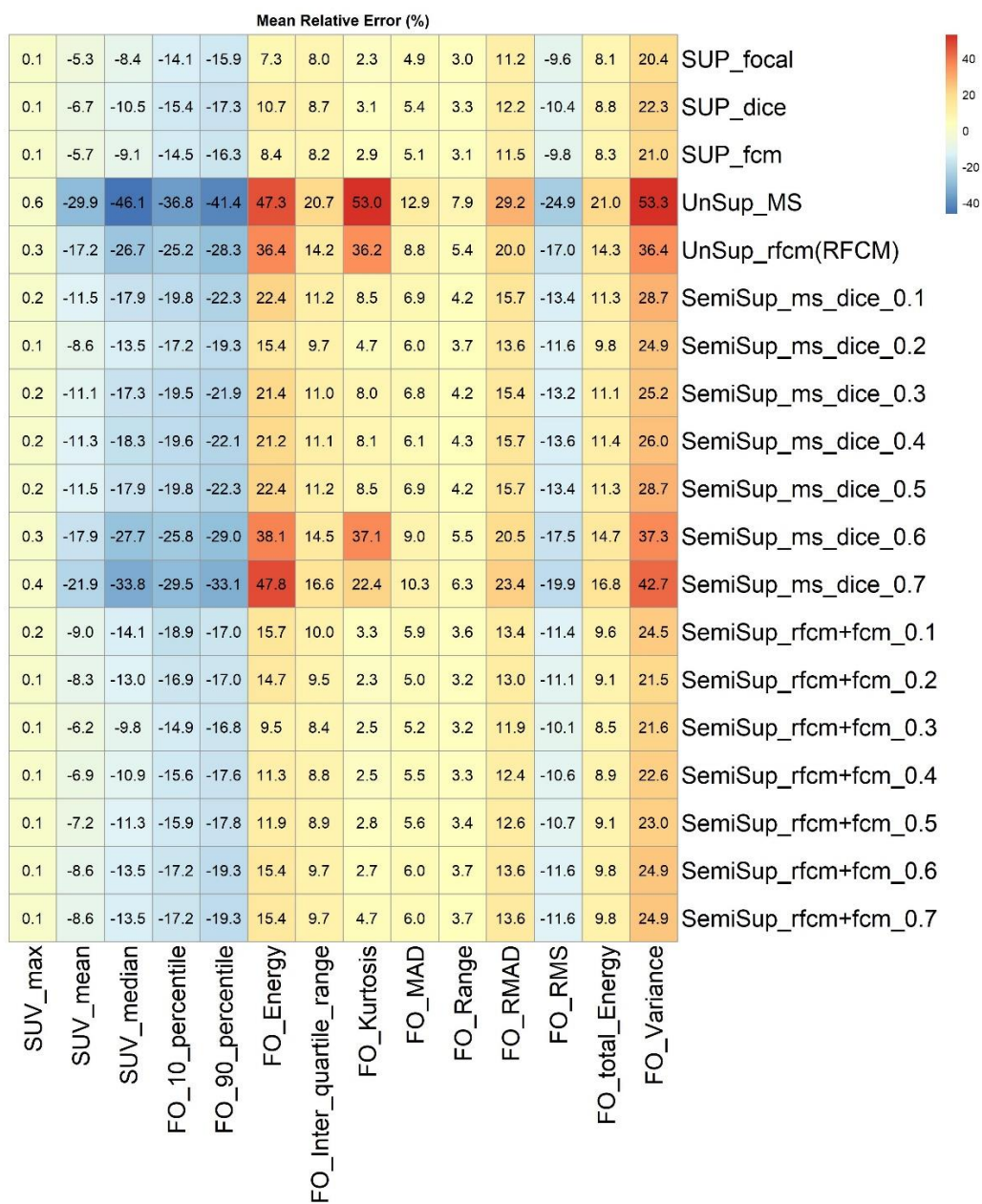

**Figure 5.** Mean relative error (MRE %) of radiomic features for the different levels of supervision and different loss functions

Figure 6 shows a few representative outliers to help you better understand our techniques. In these instances, the investigated approaches with various levels of supervision failed to properly segment the lymphoma lesions. This figure illustrates how false positives (left) and missed lesions (right) were caused by the low uptake of lesions (right) and high uptake in the background (left). The presence of nearby tissues with a relatively high uptake that could be mistaken for the tumor is another factor that can cause errors in the model predictions (see Figure 6).

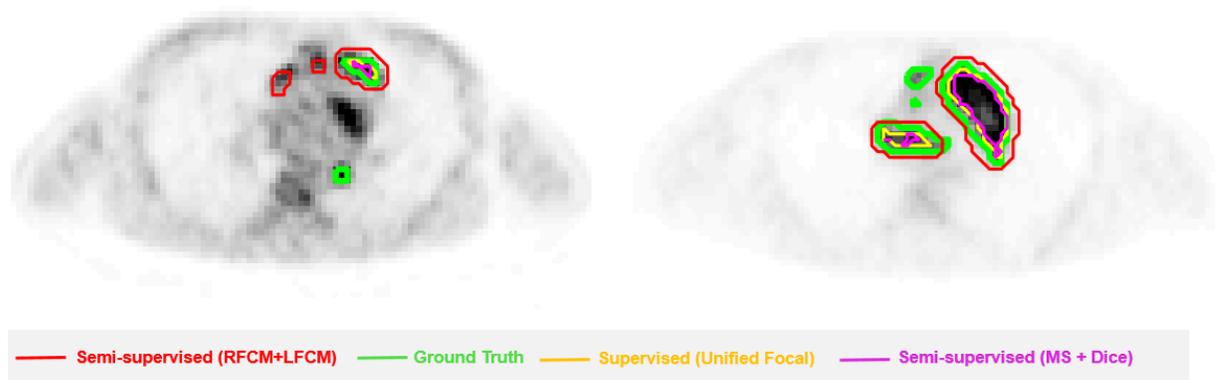

**Figure 6.** Axial views of ground truth and 3D U-net trained by different levels of supervision on different cases where failure was observed resulting in outliers.

Figure 7 shows the probability map predictions of our network by semi-supervised learning based on FCM losses (RFCM+αLFCM) superimposed on the axial slice of a PET scan. Four cases of different probability threshold settings are displayed. The range of probability maps is shown on the image for each cluster.

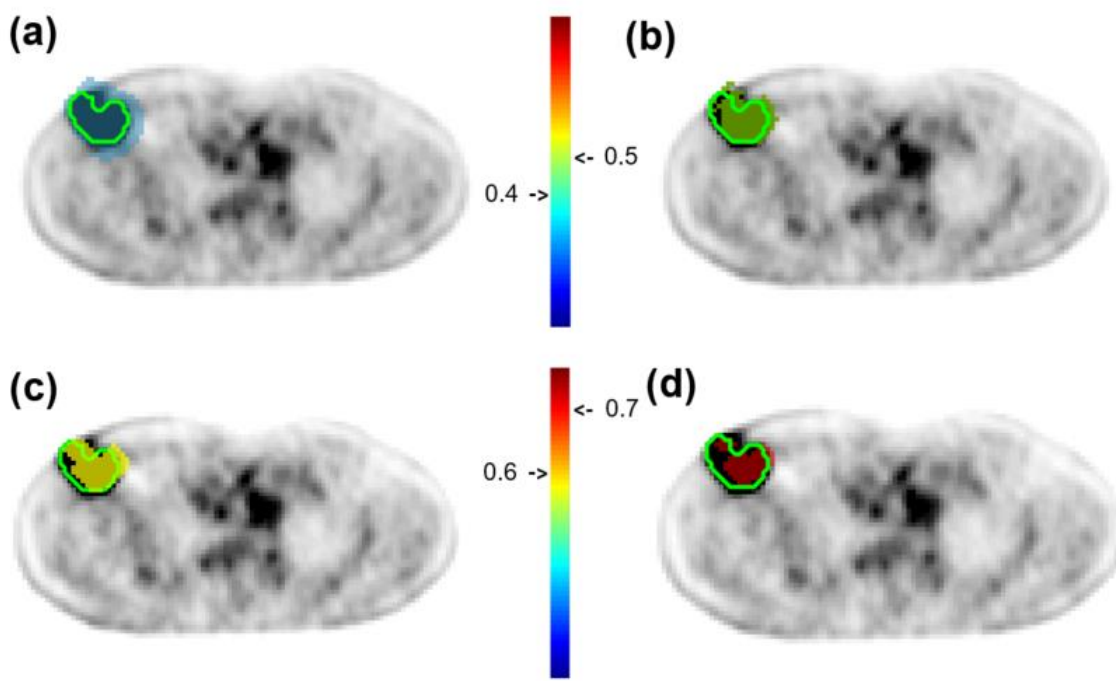

**Figure 7.** Examples of the output of the semi-supervised (RFCM+LFCM). Outcome probabilities displayed on the axial view of slices extracted from the PET 3D volume of a patient in the dataset. The green line represents the tumor area provided as ground truth, while the different colors show different areas of tumor probabilities. Four examples of customized outputs based on different probability threshold settings, as shown by the color bars on the right side of the images. In (a) only the areas containing pixels with tumor probabilities above 0.4 are displayed, above 0.5 in (b), above 0.6 in (c) and above 0.7 in (d).

In Figure 8, we considered the impact of α in $L^{\alpha}_{semi-FCM}(q = 2, \beta = 0.0016)$ and $L^{\alpha}_{semi-MS}(\eta = 10^{-6})$ on the performance of the lesion segmentation. As shown in Table 1, the performance of semi-supervised approach $L^{\alpha}_{semi-FCM}$ (RFCM+αLFCM) was consistently higher than $L^{\alpha}_{semi-MS}$ (MS+αDice) for the different α values of (0.1, 0.2, 0.3, 0.4, 0.5, 0.6, 0.7) that we considered (Figure 8).

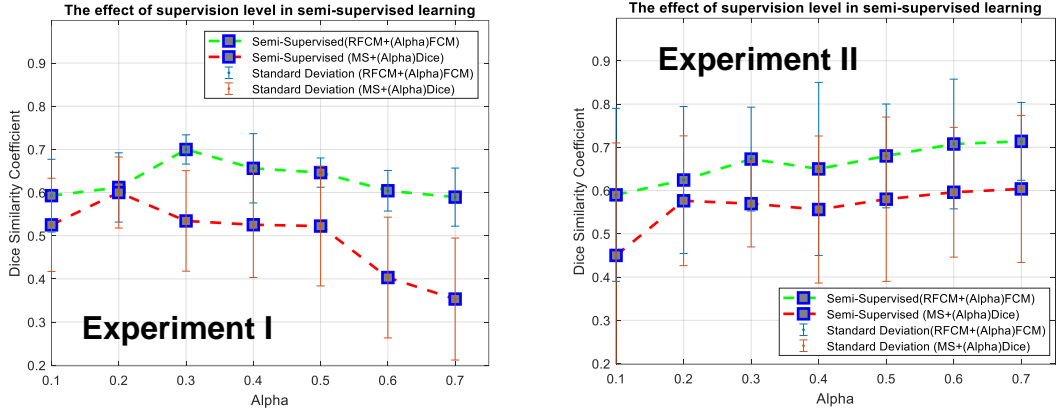

**Figure 8.** The effect of supervision level by changing α on the segmentation performance (Dice score) in two semi-supervised approaches: RFCM+αLFCM and MS+αDice for experiment I and II.

## Discussion

There are a number of challenges to reliable tumor segmentations from PET images: (i) The limited spatial resolution of PET images; (ii) Blurred boundaries between the adjacent functional regions due to partial volume effect [45] (these result in underestimated activity in small lesions [11]; (iii) Low signal-to-noise ratios in PET images due to limitations in injected radioactivity, scanner sensitivity, and scan durations. These also vary when using different systems, acquisition protocols, and reconstruction settings in a multi-center context, further challenging segmentations; (iv) The wide variability in sites, shapes, size, and heterogeneity of lesion uptakes might reduce the generalization of segmentation methods to only some specific cases. More specifically, delineation of the entire lesions in PET scans of lymphoma patients is challenging and time-consuming especially for patients with a heavy disease burden [46]. Other issues with tumor segmentation can be related to the type of cancer, e.g. proximity to physiological high uptake regions. Several techniques have been proposed to tackle the above-mentioned limitations [11,47,48]. Moreover it has been shown that high repeatability of segmentation for smaller lesions in PET is hard to achieve since partial volume effect affects the apparent tumor uptake [49].

Most conventional methods for segmentation are based on the classification of each voxel in the PET image to the tumor (or a specific normal organ) vs. background region. Fuzzy techniques such as FCM encounter the fuzziness of the tumor (lesion) edges in PET scans. However, their applications for clinical workflow are limited since they are time-consuming and need user interaction. On the other hand, to develop a generalized AI-based automated segmentation approach, there is a need to overcome the ground truth inconsistency originating 1) from inter/intra-observer variabilities and lack of gold standard approaches for segmentation, and 2) domain shift between the dataset that has been used for training and unseen data.

In this study, we applied two semi-supervised approaches for tumor segmentation in PET scans. Semi-supervised learning approaches were implemented, integrating two loss functions designed for unsupervised learning based on the FCM cost function and Mumford-Shah formulation. In addition to their inherent noise-suppressing capabilities [50] and higher accuracy for tumor segmentation tasks [11], CNNs have much shorter prediction times than traditional segmentation methods like FCM. FCM loss function is suitable for training deep networks for tumour segmentation by incorporating the classical FCM objective function since FCM has the ability to consider the fuzzy edges without supervision. FCM loss previously showed good performance for tumor segmentation in SPECT/CT images [30]. Fuzzy clustering loss function can be used as the supervised and unsupervised loss function based on the definition of the desired output in the loss and some modifications in the training process adaptively. Most of the suggested existing CNN-based unsupervised learning methods [25,51–53] usually require complex pre- and /or post-processing. In our proposed approach, the only required modification to apply the suggested approach on new datasets from different centers is the supervision level, α.

Figure 7 shows the probability map predictions of our network by semi-supervised learning based on FCM losses (RFCM+αLFCM) superimposed on the axial slice of a PET scan. Increasing the probability percentage shrinks the predicted area from around to the inside of the tumor. The parameter α allows us to regulate the degree of supervision in semi-supervised approaches. For small value of α, training the network is focused on intensity distribution characteristics of the image rather than the ground truth labels of annotated training data. We showed that combined unsupervised RFCM and supervised LFCM (RFCM+αLFCM), performed better compared to integration of unsupervised MS loss and supervised Dice loss (MS+αDice). RFCM+αLFCM with α=0.3 showed the best performance compared to the semi-supervised approach based on MS loss (p-value <0.01) with the percent relative error (RE%) of $SUV_{max}$ quantification less than 1%.

As Figure 8 shows, in experiment II, training and test data are from DLBCL cases. Besides the training data was mainly composed of data from BCC center. By increasing α (from α=0.1 to α=0.7) the impact of supervised loss was increased and the segmentation performance on test data (also from BCC) improved (Figure 8). In experiment I, the reduction in performance when we increase α (Figure 8), can be explained by domain shift. As the weight of the supervised term grows, more "domain shift" issues arise in the model; since the model was trained on data that were mostly from BCC center and the supervised learning does not generalize well on test data from SM (Figure 8). In other words, since most of the data used for training and testing are not drawn from the same distribution (center), increasing the weight of the supervised term decreases the segmentation performance. The performance drop in experiment I was higher in MS based semi-supervised approach (MS+αDice) compared to FCM based (RFCM+α LFCM). Experiment I is close to the real-world scenario that we mainly apply trained models on unseen data from external centers. While, in the case of having test data from a center with limited contribution to training data, the segmentation performance was decreased due to domain shift phenomena. On the other hand, when only annotated data with ground truth inconsistency is available, decreasing the effect of supervised term in the semi-supervised approach will emphasize the unsupervised learning from data itself.

We faced some limitations related to the limited stage and interim scans that mostly includes small size lymphoma lesions ($< 2$cm) that are challenging for segmentation even in supervised approaches. The model frequently struggled to correctly segment images with relatively small tumor regions. Some of the labeled cases from the SM center were segmented by thresholding techniques (40%) and this can increase ground truth inconsistencies and we removed them from our training data Semi-supervised approaches were implemented by combining the loss functions for unsupervised learning and the loss function defined based on the labels for supervised learning that needs the compatibility of these losses. Combining the loss functions for unsupervised learning and the label-based loss function for supervised learning, requires the compatibility of these losses, led to the implementation of semi-supervised techniques.

## Conclusion

Given the wide availability of unlabeled PET data, it is possible to leverage the need for high-quality annotated data via semi-supervised approaches. To this end, we evaluated two semi-supervised approaches for 3D segmentation of lymphoma lesions. Our study revealed that a semi-supervised approach with a well-designed loss function could be a great alternative when only having access to a limited amount of annotated data or ground truth inconsistencies. Particularly, a semi-supervised method that combines an unsupervised loss function with a supervised loss from the same category (region, boundary, or distribution) can achieve promising results. Compared to supervised approaches trained on a smaller amount of labeled data, semi-supervised learning via FCM loss (RFCM+$\alpha$LFCM) demonstrated improved performance. Semi-supervised approaches have significant potential for automated segmentation workflows due to the time-consuming nature of expert manual delineations and intra-observer variabilities that result in inconsistent ground truths.

## Acknowledgements

This research was supported by the Canadian Institutes of Health Research (CIHR) Project Grant PJT-173231, in part through computational resources and services provided by Microsoft for Health, and the Swiss National Science Foundation under grant SNRF 320030_176052.